\documentclass[pra,twocolumn,superscriptaddress,showpacs,amsmath,amssymb]{revtex4}

\usepackage{amsmath,amssymb}
\usepackage{graphicx,color}
\usepackage{booktabs}

\usepackage[latin1]{inputenc}

\newcommand{\myinputfig}[1]{\includegraphics[width=0.95\columnwidth]{#1.eps}}

\newtheorem{theorem}{Theorem}
\newtheorem{algorithm}[theorem]{Algorithm}
\newtheorem{proposition}[theorem]{Proposition}

\newcommand{\veps}{\varepsilon}
\renewcommand{\Im}{\operatorname{Im}}
\renewcommand{\Re}{\operatorname{Re}}
\newcommand{\vect}{\operatorname{vec}}

\begin{document}

\title{Computing singularities of perturbation series}

\author{Simen Kvaal} 
\email{simen.kvaal@cma.uio.no}
\affiliation{Centre of Mathematics for Applications, University of
  Oslo, N-0316 Oslo, Norway} 
\author{Elias Jarlebring}
\affiliation{Departement Computerwetenschappen, K.U. Leuven,
  Celestijnenlaan 200 A, B-3001 Heverlee, Belgium}
\author{Wim Michiels}
\affiliation{Departement Computerwetenschappen, K.U. Leuven,
  Celestijnenlaan 200 A, B-3001 Heverlee, Belgium}


\begin{abstract}
  Many properties of current \emph{ab initio} approaches to the
  quantum many-body problem, both perturbational or otherwise, are
  related to the singularity structure of Rayleigh--Schrödinger
  perturbation theory.  A numerical procedure is presented that in
  principle computes the complete set of singularities, including the
  dominant singularity which limits the radius of convergence.  The
  method approximates the singularities as eigenvalues of a certain
  generalized eigenvalue equation which is solved using iterative
  techniques. It relies on computation of the action of the perturbed
  Hamiltonian on a vector, and does not rely on the terms in the
  perturbation series. Some illustrative model problems are studied,
  including a Helium-like model with $\delta$-function interactions
  for which Møller--Plesset perturbation theory is considered and the
  radius of convergence found.  
\end{abstract}

\maketitle

\section{Introduction}
\label{sec:introduction}

Many-body perturbation theory (MBPT) has been one of the most popular
approaches for \emph{ab initio} many-body structure calculations, both
in atomic, nuclear and chemical physics. Low-order Møller--Plesset (MP)
partial sums were for many years considered highly accurate and the
method of choice for calculations of ground state energies. However,
in recent years it has become clear that the convergence properties
are not that simple, and that plain MBPT more often than not is
divergent \cite{Christiansen1996,Olsen1996,Dunning1998,Stillinger2000,Leininger2000,Roth2010}.

Divergent series in this context can still be very useful. The series
should be considered not as a final answer, but -- as a Taylor series
of a particular function with a particular singularity structure -- be
analyzed to obtain new ways of summing the series. Indeed the now
extremely popular {\em coupled cluster method}, which has to a large
extent supplanted low-order MBPT as the most effective method for
\emph{ab initio} structure calculations, can be described in terms of
summations of selected classes of diagrams (\emph{i.e.}, selected
terms in the series) to infinite order \cite{Bartlett1981}. Numerous
other ways of resumming the series give improvements of the
convergence, such as Padé or algebraic approximants \cite{Brandas1970,Goodson2000,Goodson2003,Sergeev2006,Roth2010}. Especially
in nuclear physics, summations of classes of diagrams to infinite
order, such as the random-phase approximation \cite{Siu2008}, have wide-spread use.

The performance of MBPT and the various resummation techniques is
determined by the singularity structure of the energy eigenvalue maps
$E_n(\lambda)$, where $\lambda$ is the perturbation parameter. The
determination of these singularities, which are of branch-point type,
is therefore crucial, but also very involved.  Current approaches use
the terms in the series to estimate the location of the singularities,
perhaps in combination with approximants \cite{Hunter1980,Pearce1978,Zamastil2005,Goodson2000,Goodson2003,Sergeev2006}. However,
due to a theorem by Darboux \cite{Pearce1978,Sergeev2006} the
asymptotic form of the series only gives information about the
dominant singularity, \emph{i.e.}, the one closest to the origin, and
such methods may also be sensitive to round-off errors \cite{Sergeev2006}. It
is also possible to do (very expensive) parameter sweeps of $\lambda$
to locate avoided crossings \cite{Helgaker2002,Sergeev2005,Herman2009}, thereby
discovering empirically some singularities, but only those with small
imaginary parts.

In this article, we present a general and reliable numerical procedure
for computing in principle the \emph{complete} set of singularities of
the eigenvalue maps $E_n(\lambda)$ and a procedure for determining the
dominant singularity in standard Rayleigh--Schrödinger (RS)
perturbation theory from the results, thereby finding the radius of
convergence (ROC) of the series. The method relies solely on being
able to compute the action of the Hamiltonian on a vector, which is
compatible with the common approach of using the full
configuration-interaction (FCI) methodology for computing the series
terms \cite{Laidig1985,Handy1985,Christiansen1996,Olsen1996}. We apply
the numerical procedure to several examples and discuss
the results.

We have chosen the examples for their instructive nature and the fact
that we can compare with an explicit analysis.
We analyze a simple harmonic oscillator
with a $\delta$-function 
potential added \cite{Patil2006}, a three-electron quantum wire model
in one spatial dimension \cite{Reimann2002}, and an MP treatment of a
Helium-like model with $\delta$-function interactions, which was also
considered recently in detail by Herman and Hagedorn \cite{Herman2009}
using parameter sweeps. In this paper we make conclusions about the
ROC of this model. We use only very simple basis
sets based on standard discretization techniques. The two first
examples illustrate the properties of our numerical procedure, while
the final example illustrates an application of moderate complexity.

Our method is based on the characterization of the singularities as
branch points in the complex
plane \cite{Kato1995,Schucan1973,Helgaker2002}. Those are equivalently
the points $\lambda_*$ where eigenvalues coalesce. It has been shown by the
authors \cite{Jarlebring2010} that these points can be
approximated to high precision by solving a particular two-parameter
eigenvalue problem. More precisely, we find $\lambda(\veps)$
such that a pair eigenvalues have a small relative distance $\veps$,
\emph{i.e.}, $E_n(\lambda)$ and $E_m(\lambda)=(1+\veps)E_n(\lambda)$ are
both eigenvalues. 
We adapt this result, exploit the structure of the Hamiltonian
matrix, and combine this with modern solvers for eigenvalue
problems. 

The dominant singularity for RS perturbation theory for the ground
state is the branch point $\lambda_{*,0}$ closest to the origin where
$E_0$ meets $E_n$, $n\neq 0$ \cite{Schucan1973,Kato1995}. The second
part of the numerical method is a procedure that tracks the
eigenvalue branches from the branch points $\lambda_{*}$ to the
origin, thereby determining if it is the dominant branch point
$\lambda_{*,0}$. We have chosen to focus on locating the dominant
branch point since the ROC is a fundamental property of the
perturbation series. The tracking procedure can equally well be applied to
study other branch points.

After discussing the numerical method in Section~\ref{sec:method}, we
apply it to the model problems in Section~\ref{sec:results}. Finally
we present our conclusions in Section~\ref{sec:conclusion}.

\section{Method}
\label{sec:method}

\subsection{Properties of RS perturbation series}
\label{sec:perturbation-theory}

Consider a Hamiltonian matrix
$H$ of dimension $N$ on the form
\begin{equation*}
  H(\lambda) = H_0 + \lambda V,
\end{equation*}
where $V$ is treated as a perturbation, and where $\lambda$ is a
complex parameter introduced for convenience. For the actual physical
system we have $\lambda = \lambda_\text{phys} \in \mathbb{R}$. Both
$H_0$ and $V$ are Hermitian matrices. The eigenvalues $E_n(\lambda)$
of $H(\lambda)$ are the $N$ roots of the characteristic polynomial
$\det[H(\lambda) - EI]$, where $I$ is the identity matrix. The
eigenvalues $E_n(\lambda)$ are the branches of an $N$-valued algebraic
function, whose only singular points (denoted $\lambda_*$) are in fact
of branch-point type \cite{Kato1995,Schucan1973,Helgaker2002}.

For Hermitian matrices the branch-points come in complex conjugate
pairs. There are no real branch points, and in the generic case (see
Section~\ref{sec:genericicity}) all
branch points are of square-root type. For
sufficiently small $\lambda-\lambda_*$ the eigenvalues can be expanded
in a Puiseux series around each branch point \cite{Schucan1973}. This is
contained in Katz' theorem \cite{Katz1962} which we state here:
\begin{theorem}
  Suppose $H(\lambda)=H_0 + \lambda V$ is generic in the sense that $H_0$ and $V$ are
  Hermitian and chosen at random. Then for any pair of branches $E_n$
  and $E_m$ there exists 
  a branch point $\lambda_*$ at which $E_n(\lambda_*) =
  E_m(\lambda_*) = b_{nm}$. Moreover, for sufficiently small
  $\lambda-\lambda_*$ there exists a constant $c_{nm}$ such that
  \begin{equation*}
    E_n(\lambda) = b_{nm} + c_{nm} (\lambda - \lambda_*)^{1/2} + \mathcal{O}(\lambda-\lambda_*)
  \end{equation*}
and
  \begin{equation*}
    E_m(\lambda) = b_{nm} - c_{nm} (\lambda - \lambda_*)^{1/2} + \mathcal{O}(\lambda-\lambda_*),
  \end{equation*}
where it is to be understood that the same branch of the square-root
function is to be used in both equations.
\end{theorem}
Katz' theorem may be viewed as a generalization of the well-known
Wigner--von Neumann non-crossing rule \cite{Schucan1973}. It is interesting that
\emph{all} eigenvalue pairs are involved at \emph{some} branch point,
which implies that the function $E_0(\lambda)$ actually can be
analytically continued to \emph{any} excited state $E_n(\lambda)$.

Finite-dimensional Hamiltonians usually arise due to some
discretization in form of a finite basis set, \emph{e.g.}, using the FCI
methodology. The singularity structure of the full problem is richer
than in the finite-dimensional case, but we postpone a brief
discussion to Section~\ref{sec:general-discussion}. 

In RS perturbation theory for the ground state one computes a
truncated Taylor series for $E_0(\lambda)$, viz,
\begin{equation*}
  E_0(\lambda) = \sum_{k=0}^K E_{0,k}\lambda^k + \mathcal{O}(\lambda^{K+1}), 
\end{equation*}
which is an asymptotic series approximating $E_0(\lambda)$ as
$\lambda\rightarrow 0$. The coefficients $E_{0,k}$ can be generated recursively by
insertion into the eigenvalue problem for $H(\lambda)$ which gives a
series usually represented in form of Feynman diagrams. The actual
computation of the terms become increasingly complicated for
higher-order terms for many-body systems (in practice, one rarely
computes more than sixth-order series using diagrammatic techniques), but if
$H(\lambda)$ is is available as a matrix or as a procedure that
computes matrix-vector products, the high-order terms are straightforward to
compute \cite{Laidig1985,Roth2010}.

One of the important questions we consider in this paper is whether
the truncated series is
convergent as $K\rightarrow\infty$ for 
$\lambda=\lambda_\text{phys}$, that is to say whether the ROC is greater than
$\lambda_\text{phys}$ or not. As a Taylor series, the ROC is given by
$|\lambda_{*,0}|$, where $\lambda_{*,0}$ is the smallest branch point,
called the dominant branch point, where the branch belonging to
$E_0(0)$ meets a different branch $E_n$, $n\neq 0$. We say that the
$E_0$ and $E_n$ branch at $\lambda_{*,0}$ \cite{Kato1995,Baker1971,Schucan1973,Goodson2004}. We remark that in
other perturbation theories, like the folded diagram series for the
effective interaction in nuclear physics \cite{Schucan1973}, other
branch points may be dominant.

Thus, to compute the dominant singularity we are looking for the
points $\lambda_*\in\mathbb{C}$ \emph{not} on the real line such that
$E_n(\lambda_*) = E_0(\lambda_*)$ for $n\neq 0$. The ROC is then $R =
\min\{|\lambda_*|\} = |\lambda_{*,0}|$. Instrumental to this we
consider the values of $\lambda$ such that the matrix $H(\lambda) =
H_0 + \lambda V$ has a double eigenvalue. In what follows we call such
a value of $\lambda$ a \emph{critical value}. It usually is a branch
point, but in some cases we also get spurious solutions.

\label{sec:doubleeig}

\subsection{Computing the $m$ smallest critical values} 
\label{sec:compm}
We saw above that it is possible to characterize the singularities of a
perturbation series by computing  $\lambda$ such that $H(\lambda)$ has
a double eigenvalue.  
The problem of finding all $\lambda$ such that a matrix depending
linearly on $\lambda$ has a double eigenvalue has been considered by
the authors elsewhere \cite{Jarlebring2010}. The derivation of the method
presented here is based on  
a result in Ref.\ \cite{Jarlebring2010} stating that  
all solutions can be approximated by the 
solutions of a generalized eigenvalue problem defined as follows. We let $\veps>0$ be a
small scalar, called the regularization parameter, and define the matrices $\Delta_0(\veps),\,
\Delta_1(\veps)\in\mathbb{C}^{N^2\times N^2}$  by
\begin{equation*}
\Delta_0(\veps) = -I \otimes V+(1+\veps) V\otimes
I 
\end{equation*}
and
\begin{equation*}
\Delta_1(\veps) = I\otimes H_0-(1+\veps)H_0\otimes I, 
\end{equation*}
where $\otimes$ denotes the Kronecker product. 
Consider now the generalized eigenvalue problem
\begin{equation}
\lambda \Delta_0(\veps)v=\Delta_1(\veps)v,\ \ \ \label{eq:deltaevp}
\end{equation}
where $v = v_1\otimes v_2$.
An important result is is that the solutions of
\eqref{eq:deltaevp} approximate all $\lambda$ such that $H(\lambda)$
has a double eigenvalue.

Approximations of the eigenvalues $E_n(\tilde\lambda)$ can be obtained
from the eigenvalue problem for the matrix $H_0+\tilde\lambda V$,
where $\tilde\lambda$ satisfies (\ref{eq:deltaevp}). To shed a light
on this approach, one can prove that the eigenvalues of
(\ref{eq:deltaevp}) correspond to the set of values of $\lambda$ for
which the matrix $H_0+\lambda V$ has two eigenvalues within a relative
distance of $\veps$, \emph{i.e.}, a pair of eigenvalues of the form $E$ and
$E(1+\veps)$. This is clearly a relaxation of the problem. A suitable
choice of $\veps$, as well as other implementation aspects, have been
studied in detail \cite{Jarlebring2010}. This includes the exclusion of 
spurious solutions. It is also showed that the error in
$\tilde\lambda$ behaves like $\mathcal{O}(\veps^2)$.

Although the above method allows us to compute all critical values of
$\lambda$, it may be computationally prohibitive for large problems. 
The computational complexity is determined by the solution of the
generalized eigenvalue problem (\ref{eq:deltaevp}), which requires
$\mathcal{O}(N^6)$ operations if all eigenvalues are computed with a
general purpose  method. To overcome this problem we use an
iterative method known as the Arnoldi method \cite{Saad1992},
to compute the $m$ smallest eigenvalues of (\ref{eq:deltaevp}), where
$m$ is a given integer.  

The Arnoldi method generalizes the familiar Lanczos iterations
employed in FCI calculations to non-Hermitian matrices, and only
requires an efficient computation of the matrix-vector product
associated with the eigenvalue problem. The matrix-vector product
associated with \eqref{eq:deltaevp} is
\begin{equation}\label{eq:iterationmatrix}
y=\Delta_1(\veps)^{-1}\Delta_0(\veps)x.
\end{equation}
Let $X,Y\in\mathbb{C}^{N\times N}$ be such that $x=\vect(X)$ and
$y=\vect(Y)$, where $\vect:\mathbb{C}^{N\times N}\rightarrow \mathbb{C}^{N^2}$ denotes the vectorization operation, \emph{i.e.},
stacking the columns of the matrix on top of each other. A key to the
success of our method is that we can 
express the matrix-vector product \eqref{eq:iterationmatrix} in terms
of the matrices $X$ and $Y$. By straightforward manipulations 
using the rules of the Kronecker product we  obtain
\begin{equation}
H_0 Y-(1+\veps) Y H_0^T=-V X+(1+\veps)XV.\label{eq:DeltaDeltaMat}
\end{equation}
This matrix equation, where $Y$ is the unknown, is a matrix equation
known as a {\em Sylvester equation}.  The right-hand side can be evaluated 
in $\mathcal{O}(N^3)$ operations.
The Sylvester equation can be solved in
$\mathcal{O}(N^3)$ operations by using the Bartels-Stewart  
algorithm \cite{Bartels1972}, which is a standard method for Sylvester
equations. Hence, by exploiting the structure in this way, the matrix-vector
products of \eqref{eq:deltaevp} can be efficiently computed 
in  $\mathcal{O}(N^3)$ operations. If $H_0$ is diagonal, this can be
improved to $\mathcal{O}(N^2)$ operations, which is seen as follows.

Let $Y=:(y_1,\ldots, y_N)$ and
$(c_1,\ldots,c_N)$ the columns of the right-hand side of
\eqref{eq:DeltaDeltaMat}. Suppose the diagonal entries of $H_0$ are
$H_{0,i,i}$,\; $i=1,\ldots,N$. 
It is straighforward to show that column $i$ of 
\eqref{eq:DeltaDeltaMat} can be written as the solution of a linear
system with a diagonal matrix, 
\begin{equation}
 y_i= \operatorname{diag}(d_1,\ldots,d_N)
c_i\label{eq:diaglinsys}
\end{equation}
where 
\[
d_j = \frac1{H_{0,j,j}-(1+\veps)H_{0,i,i}},\;j=1,\ldots,N.
\]
Now note that we can compute a column vector of $Y$ using
\eqref{eq:diaglinsys} with only $\mathcal{O}(N)$ operations. Hence, the 
Sylvester equation corresponding to diagonal $H_0$ can be solved in $\mathcal{O}(N^2)$ operations. In
the simulations in Section~\ref{sec:wire} we will use this approach,
whereas using the Bartels-Stewart algorithm \cite{Bartels1972} turned
out to be more robust in Section~\ref{sec:helium}.

Two additional properties of the Arnoldi method makes it particularly
suitable for our purposes.
\begin{itemize}
\item
    As we shall illustrate in  Section~\ref{sec:results} the
    overall algorithm for computing the dominant branch point, outlined in
    Section~\ref{sec:parsmallest}, typically requires only a small number of
    critical values $m\ll N$. The Arnoldi method can be useful if
    not all eigenvalues need to be computed.
\item If the chosen $m$ is deemed insufficient, we wish to continue
  the iteration. The Arnoldi method can be easily resumed if more
  eigenvalues are needed. 
\end{itemize}

The procedure above describes a method which can be used for quite
large systems since the complexity of the matrix-vector product is
only $\mathcal{O}(N^2)$. The matrices $V$ and $H_0$ stem from 
discretizations and we wish to  be able to solve as large systems as
possible. We will now use that for large problems (fine
discretization/large basis set), a
somewhat accurate guess is available by solving a corresponding
smaller problem (coarser discretization/small basis set).

{\em Inverse iteration} \cite{Saad1992} is a method to compute one eigenpair
where a reasonable approximation of
the eigenvalue is already available. Inverse iteration is, similar to the
Arnoldi method, also only based on matrix vector products. It however,
does not involve any orthogonalization step.  Since it is only based
on matrix-vector products
we can use \eqref{eq:DeltaDeltaMat} directly with  inverse
iteration. 

By using the Arnoldi method for a coarse discretization, and inverse
iteration for a finer discretization, we can solve very large problems
in a reliable way in a multi-level fashion.

\subsection{Computing the dominant branch point}
\label{sec:parsmallest}

The value $\lambda_{*,0}\in\mathbb{C}$ is  the first branch point of $E_0(\lambda)$.
Since $H(\lambda_{*,0})$ has a double
eigenvalue, we can compute candidates for $\lambda_{*,0}$, \emph{i.e.} the
critical values, with the 
procedure described in Section~\ref{sec:compm}. It now remains to
determine which one of 
the candidate solutions computed with the method in
Section~\ref{sec:compm}  corresponds to~$\lambda_{*,0}$. 

We will use a computational approach based on following paths from
the candidate solution $\lambda$ to the origin. It is justified by the
following technical result.

\begin{proposition}\label{mainprop}
Consider a critical value $\tilde{\lambda}$ such that $|\tilde{\lambda}|<|\lambda_{*,0}|$.
Let $p:[0,1]\rightarrow \mathbb{C}$ be a parametrization of a curve
from $p(0)=\tilde{\lambda}$ to $p(1)=0$ such that
$|p(\theta)|\le|\tilde{\lambda}|$ for $\theta\in [0,1]$.  Assume that
$p$ does not pass directly through another critical value.  Then two
continuous eigenvalue functions $[0,\ 1]\ni\theta\mapsto E_n(\theta)$
and $[0,\ 1]\ni\theta\mapsto E_b(\theta)$ satisfying $E_a(\theta),
E_m(\theta)\in\sigma(H_0+ p(\theta)V)$ for $\theta\in[0,\ 1]$ and
$E_n(0)=E_m(0)$ are uniquely defined. Moreover, we have
\begin{equation*}
E_n(1)\neq E_0(0) \textrm{ and }E_m(1)\neq
E_0(0).
\end{equation*}
\end{proposition}

\noindent\textbf{Proof.} The first statement follows from Rouch\'e's
Theorem \cite{Krantz1999}. The second statement can be proven by
contradiction. More precisely, the statement $E_n(1)=E_0(0)$ or
$E_m(1)=E_0(0)$ contradicts with the assumption $|\tilde\lambda|<
|\lambda_{*,0}|$. \hfill $\Box$

\smallskip

From Proposition~\ref{mainprop} it follows that $\lambda_{*,0}$ is the
\emph{smallest} branch point for which one of the
corresponding curves, $E_n$ or $E_m$, terminates at $E_0(0)$. Only
the branch points with positive (or negative) imaginary parts are relevant and
some may be spurious. Denoting
all relevant numerical branch points by $\left\{\lambda_k\right\}_{k=1}^{N'}$, where
\[
|\lambda_1|\leq |\lambda_2|\leq \ldots \leq |\lambda_{N'}|,
\]
this brings us to the following algorithm.
\begin{algorithm} (Computation of the ROC)
  \label{alg:roc}
\begin{itemize}
\item[1.] Compute $E_0(0)$, set $i=1$.
\item[2.] Consider $\lambda_i$ and continue the two corresponding
  branches $E_n$ and $E_m$ for $\theta\in[0,\ 1]$, \emph{i.e.} from
  $\lambda=\lambda_i$ to $\lambda=0$. 
\item[3.] If one of the branches terminates at $E_0(0)$, then stop\\
      else
      $i=i+1$, go to step 2.
\item[5.] $R=|\lambda_i|$.
\end{itemize}
\end{algorithm}

We conclude this section with some implementation aspects. For step
2.~critical values of the parameter $\lambda$ are needed in increasing
magnitude. These can be computed by the Arnoldi algorithm as described
in Section~\ref{sec:compm}.  The value of $m$ is fixed before the iteration
starts. If it turns out to be insufficient, the Arnoldi process can
still be resumed as also outlined in Section~\ref{sec:compm}.
 
 For the continuation process in step 3.~we assume that curve $p$ is linear, \emph{i.e.}~the curve corresponding to $\lambda_i$ satisfies
 \[
 p(\theta)=(1-\theta)\lambda_i.
 \]
 As the critical values are isolated points, this line does not contain
 other critical values, with probability one.  
For the continuation of the eigenvalues we follow the eigenvalues by
sampling the line between $\theta\in [0,1]$ with sufficiently many
points. In our applications, $21$ sampling points were sufficient to follow the
eigenvalues accurately and not dominate the computation time. 

Although we have chosen to to so in our implementation, it is not
necessary to compute the whole set of 
eigenvalues along $p(\theta)$. Standard continuation techniques may
instead be used, where continuity of the eigenvalues with respect to
$\theta$ is exploited. See for example the book by Seydel \cite{Seydel2010}.

\subsection{A comment on genericity}
\label{sec:genericicity}

As stated in Section \ref{sec:perturbation-theory} the statements
concerning the nature and location of the branch points $\lambda_*$ of
the eigenvalue maps depend on the fact that $H(\lambda)$ is generic: a
statement indicating that the matrix is a ``typical'' Hermitian
matrix. Statements about generic matrices hold with probability $1$
when the matrix is chosen at random.

On the other hand, Hamiltonians are rarely generic: symmetries such as
angular momentum conservation or parity invariance lead to a natural
block structure in $H(\lambda)$ so that the eigenvalue problem
decouples into smaller, unrelated problems. It is easy to see
that the non-crossing rule may be violated, and consequently
that not all critical points of $H(\lambda)$ are branch
points if not all symmetries are removed from the system. The
numerical procedure then yields spurious real or complex solutions
corresponding to violations of the non-crossing rule or the
square-root branch point classification, respectively.

\section{Numerical results}
\label{sec:results}

\subsection{Branch points in complete basis limit}
\label{sec:general-discussion}

In this section we apply the numerical procedure to three model problems: a simple
one-dimensional harmonic oscillator with a $\delta$-function spike,
which is exactly solvable \cite{Patil2006} and equivalent to a
center-of-mass frame formulation of a parabolic two-electron quantum
wire with $\delta$-function interactions, a three-electron parabolic
quantum wire with smoothed Coulomb interactions \cite{Reimann2002},
and a helium-like model with $\delta$-function nuclear and
inter-electron interactions \cite{Herman2009}. In the latter example
we consider Møller--Plesset perturbation theory, while in the other
examples we let the perturbation $V$ be the bare inter-particle
interactions.

The Hamiltonian matrix is like in most MBPT approaches an
approximation of a partial differential operator $\mathcal{H}$
obtained by a finite basis expansion with discretization parameter
$h$, \emph{i.e.}, as $h\rightarrow 0$, the dimension $N\rightarrow \infty$
and the discrete spectrum approaches the exact limit under mild
conditions. (Special care has to be taken for the continuum spectrum
of $\mathcal{H}$ if it exists.)

One may characterize the singularities of the eigenvalue map
of $\mathcal{H}$ as $\alpha$ or $\beta$ singularities \cite{Goodson2003}. The $\alpha$ singularities are complex-conjugate
pairs of branch points with non-zero imaginary parts. These are also
called ``intruder states'' \cite{Helgaker2002}. A finite-dimensional
Hamiltonian only has $\alpha$ type branch points. The $\beta$
singularities are \emph{real} branch points corresponding to a
coalescence of an eigenvalue with the continuous spectrum. Baker
argued this is a generic feature of unconfined fermion systems \cite{Baker1971,Goodson2004,Herman2009}. The approximate Hamiltonian
will, as long as it contains sufficiently good approximations to
continuum states, have a cluster of branch points near a $\beta$
singularity. As $h\rightarrow 0$, assuming that the basis set is in
fact complete, the continuous spectrum is ``filled
out'' with discrete points, and hence there will be many close
crossings clustering around (but never equal to) a real value.

To interpret and classify the numerically found $\lambda_*$ we must
consider the non-trivial limit $h\rightarrow 0$.
In general, three cases can be expected: 
\begin{enumerate}
\item The branch point approaches a finite, complex value, and
  represents an $\alpha$ singularity.
\item The branch point approaches infinity, in case of which a
  singularity of the perturbation series disappears for the
  \emph{exact} Hamiltonian.
\item The branch point approaches a finite \emph{real}
  value. This can happen in two separate ways: {\it(i)} Since the branch
  points come in complex conjugate pairs this means that the limit
  actually becomes an analytic point as $h\rightarrow 0$, \emph{i.e.}, a
  violation of the non-crossing rule (which does not hold in the
  infinite-dimensional case). {\it(ii)} The real limit corresponds
  to a $\beta$ singularity. In that case infinitely many branch points
  must approach the same real value as $h\rightarrow 0$
  ($N\rightarrow\infty$).
\end{enumerate}

\subsection{Harmonic oscillator with $\delta$-function}
\label{sec:ho-delta}

We consider the toy model Hamiltonian \cite{Patil2006}
\begin{equation*}
  \mathcal{H}(\lambda) = -\frac{1}{2}\frac{\partial^2}{\partial x^2} +
  \frac{1}{2}x^2 + \lambda \delta(x).
\end{equation*}
 Any fixed $\lambda =
\lambda_\text{phys}$ may be taken as the actual, physical value for the toy model.

\begin{figure}[ht]
  \begin{center}
    \myinputfig{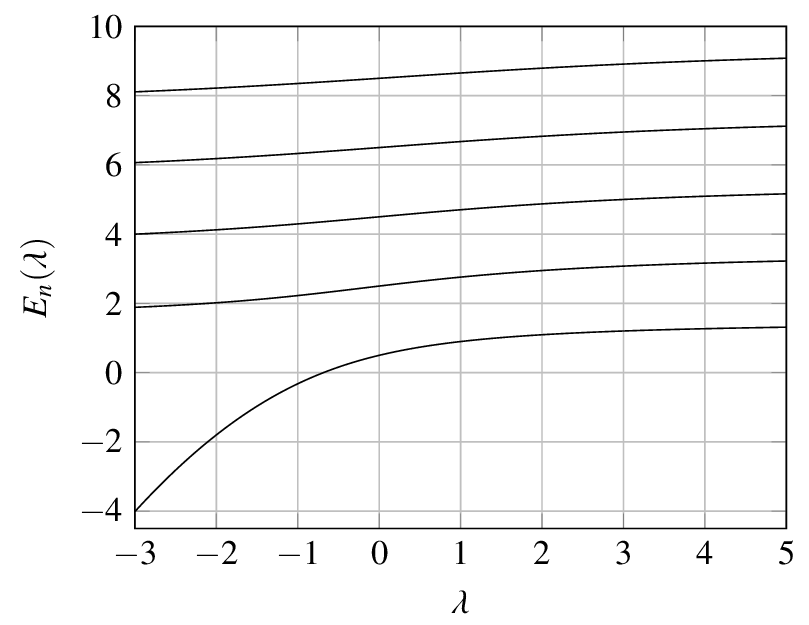}
  \end{center}
  \caption{Eigenvalues of the harmonic oscillator with a
    $\delta$-function. Only eigenvalues of even eigenfunctions are
    shown. Notice the crossing with the real axis around
    $\lambda\approx-0.6758$, which will give rise to a spurious
    solution in the numerical method.%
    \label{fig:patil-exact}} 
\end{figure}

The eigenvalue problem $(\mathcal{H}-E)\psi(x)=0$ may be solved
to arbitrary precision and represents a particularly instructive test-case for our
numerical procedure. Parity symmetry allows us to focus on even eigenfunctions
which includes the ground state. (The odd eigenfunctions are in fact
trivial since $\delta(x)\psi(x)=0$ in this case.) Figure~\ref{fig:patil-exact} shows the eigenvalues of the even eigenfunctions
as $\lambda$ is varied.

Introducing the even-numbered harmonic oscillator basis functions
$u_n(x) = \phi_{2n}(x)$ we obtain $(H_0)_{nm} = (2n + 1/2)\delta_{nm}$ and $V_{nm}
= \phi_{2n}(0)\phi_{2m}(0)$, the latter being a rank 1 matrix. Here,
\begin{equation}
  \phi_n(x) = (2^n n! \sqrt{\pi})^{-1/2} H_n(x) e^{-x^2/2},
  \label{eq:hermite}
\end{equation}
with $H_n(x)$ being the standard Hermite polynomials. It has been
shown \cite{Jarlebring2010} that the numerical procedure will 
give spurious solutions of large magnitude since $V$ has rank one; in
this case $|\lambda_*| \sim 10^{12}$. Also, one false real value
arises for $\lambda$ such that $H(\lambda)$ has a zero eigenvalue, see
Figure~\ref{fig:patil-exact}. Note that all spurious solutions are
easily detected.

\begin{figure}[ht]
  \begin{center}
    \myinputfig{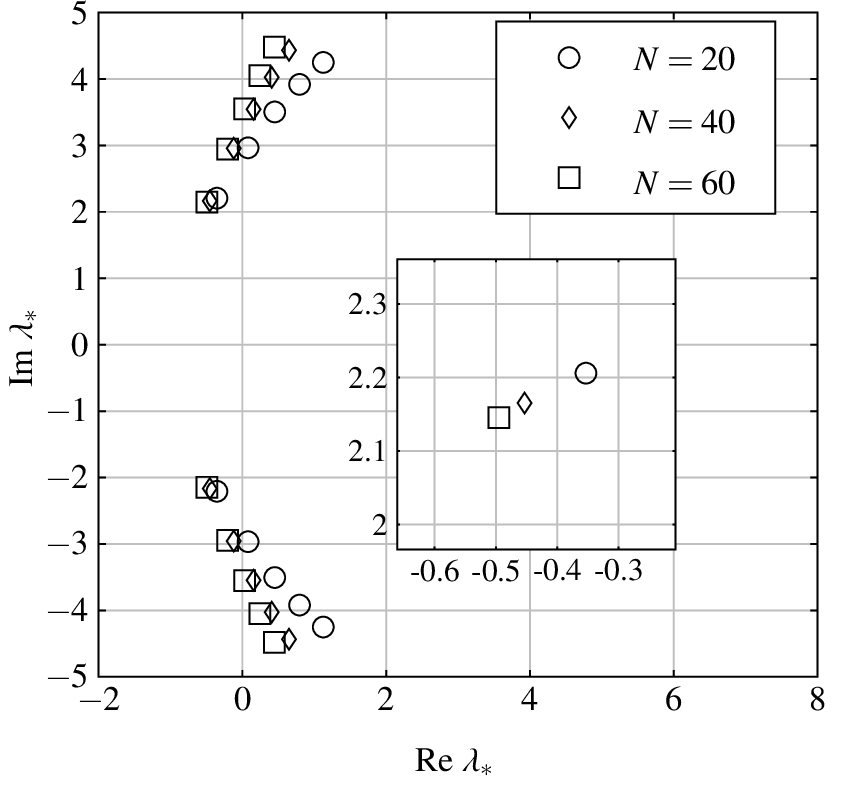}
  \end{center}
  \caption{The smallest branch points for the harmonic oscillator
    with a $\delta$-function, computed for various matrix sizes
    $N$. The dominating branch point $\lambda_{*,0}$ is shown inset.%
\label{fig:patil-model}} 
\end{figure}

Figure~\ref{fig:patil-model} shows the smallest computed branch points
for various $N$ with the dominating $\lambda_{*,0}$ inset. The results
for the various $N$ indicate that the qualitative distribution of
branch points does not change much with the basis size. For
$N\rightarrow\infty$ we then estimate that RS perturbation theory will
converge for all $|\lambda| < 2$.

We remark that it is not easy to find the branch points by doing a
parameter sweep. Figure~\ref{fig:patil-exact} does not reveal clear
avoided crossings involving any pairs of eigenvalues, which is
explained by the large imaginary parts of the various $\lambda_*$.

\begin{figure}
  \begin{center}
    \myinputfig{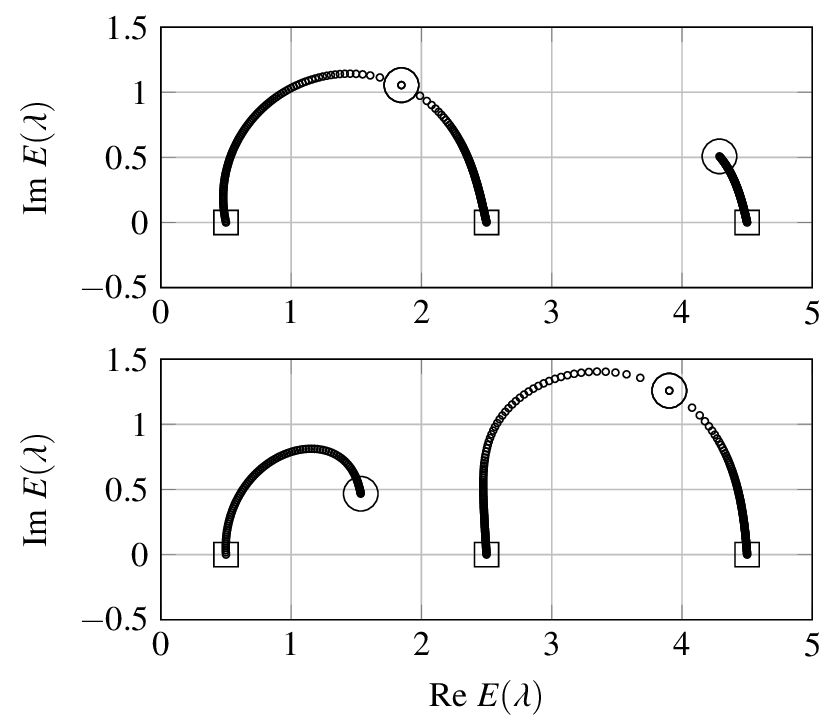}
  \end{center}
  \caption{Eigenvalue branch paths as $\lambda$ is gradually decreased
    from $\lambda_*$ to zero for two branch points of the harmonic
    oscillator with $\delta$-potential. Three eigenvalues
    are shown. The upper panel shows the paths for the dominant branch
    point $\lambda_{*,0}$, while the lower panel shows the paths for
    the first non-dominant branch point.\label{fig:ho-delta-paths}}
\end{figure}

We conclude this subsection by plotting the paths the eigenvalues trace
out when $\lambda$ is gradually decreased from $\lambda_*$ to zero, \emph{i.e.}, we consider
$\lambda(\theta)=(1-\theta)\lambda_*$ and plot eigenvalues as function of
$\theta$. Figure~\ref{fig:ho-delta-paths} shows the result for
$\lambda_{*,0}$ and one other branch point. This illustrates the
continuation process described in Section~\ref{sec:parsmallest}.

\subsection{Three-electron quantum wire}
\label{sec:wire}

The next numerical calculation is on a one-dimensional model of a
three-electron parabolic quantum wire, called so due to the
quasi-one-dimensional confinement \cite{Reimann2002}. The electrons
interact via a regularized Coulomb potential of the form
\begin{equation*}
  u(x_1,x_2) \propto \frac{1}{\sqrt{|x_1-x_2|^2 + a^2}},
\end{equation*}
where in our calculations we have set $a = 0.1$. The Hamiltonian is then of the form
\begin{equation*}
  \mathcal{H}(\lambda) = \sum_{i=1}^3
  \left(-\frac{1}{2}\frac{\partial^2}{\partial x_i^2} +
    \frac{1}{2}x^2_i \right) + \lambda \frac{1}{2}\sum_{i\neq j} u(x_i,x_j),
\end{equation*}
where we have introduced the parameter $\lambda$ which, in the chosen
units, measures the relative strengths of the interactions compared to
the semiconductor bulk and the size of the trap. Again, any fixed
$\lambda=\lambda_\text{phys}$ can be taken to be the actual value.

Due to the harmonic confinement, the spectrum of $\mathcal{H}(\lambda)$ is
discrete for all $\lambda$. It can be shown using a theorem
due to Kato \cite{Kato1995} that for all complex $\lambda$ \emph{all}
the eigenvalues depend analytically on $\lambda$, \emph{i.e.}, there are no
singularities at all. This is basically due to the boundedness of
$u(x_1,x_2)$. Thus the ROC is infinite in the exact
problem, and any perturbation approach should converge. A
discretization will, however, necessarily produce branch 
point singularities, which will approach infinity or real values as the
discretization is made finer.

We use a standard discretization based on Slater determinants
constructed from spin-orbitals on the form $\phi_n(x)\chi_\sigma(s)$,
where $\phi_n(x)$ are the harmonic oscillator functions
\eqref{eq:hermite} and $\chi_{\pm 1/2}$ are the spinor basis
functions. For a given $M$ we use all possible determinants
created from spin-orbitals with $\sum_{i=1}^3 n_i \leq M$. Thus, we
include all unperturbed three-body harmonic oscillator states of
energy less than $M + 3/2$. We restrict our attention to the lowest
possible total spin projection $S_z = \frac{1}{2}$.

This yields matrices $H_0$ and $V$ of dimension $N=O(M^2)$ when we
separate out the center-of-mass motion which is a dynamical symmetry --
the center-of-mass moves like a free particle in a harmonic
oscillator. We only consider even-parity wave functions, which
includes the ground state. These are the only symmetries of the
Hamiltonian operator, resulting in matrices for which the generic
statements hold.

Having obtained these matrices, we compute the branch points
$\lambda_*$ for $\veps = 10^{-4}$ and also deduce the ROC using our
numerical procedure. It is worthwhile to mention that in this case,
the tracking procedure reveals that the dominant singularity is a
critical value far from being the smallest.

\begin{figure}[ht]
  \begin{center}
    \myinputfig{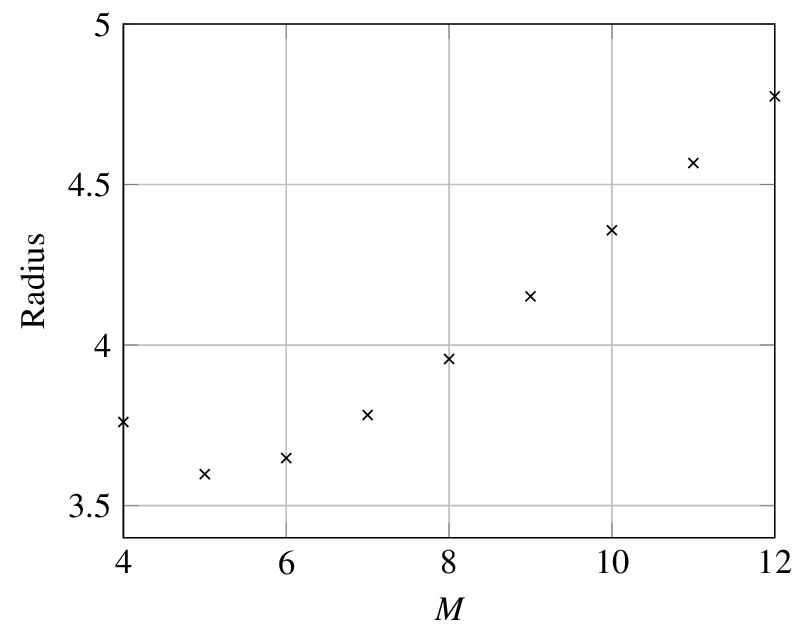}
    \caption{The radius of convergence as function of the number of
      oscillator shells $M$ for the quantum wire model. A clear linear
      tendency towards $R=\infty$ is shown.\label{fig:qdot-ROC}}
  \end{center}
\end{figure}

It is instructive to study the behavior of the ROC as function of
$M$, as shown in Figure~\ref{fig:qdot-ROC}. As expected it seems to
approach infinity linearly with $M$ as the discretization becomes
finer.

\begin{figure}[ht]
  \begin{center}
    \myinputfig{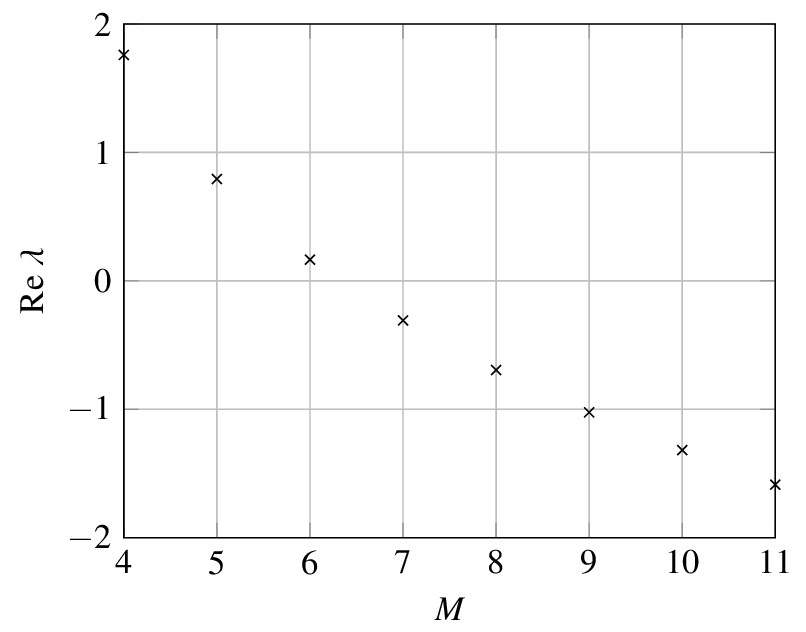}
    \caption{The real part of the dominant singularity $\Re\lambda_{*,0}$
      as function of $M$ for the quantum wire model. It changes sign around
      $M=7$, so that $\lambda_{*,0}$ changes from a back-door to a
      front-door singularity.\label{fig:qdot-intruder}}
  \end{center}
\end{figure}

If one tries to characterize $\lambda_{*,0}$ as an intruder state, it
is revealed in Figure~\ref{fig:qdot-intruder} that $\Re\lambda_{*,0}$
changes sign as $M$ is increased. This means that the characterization
as ``front-door'' or ``back-door
intruders'' \cite{Helgaker2002} is dependent on the basis used.

\begin{figure}[ht]
\begin{center}
\myinputfig{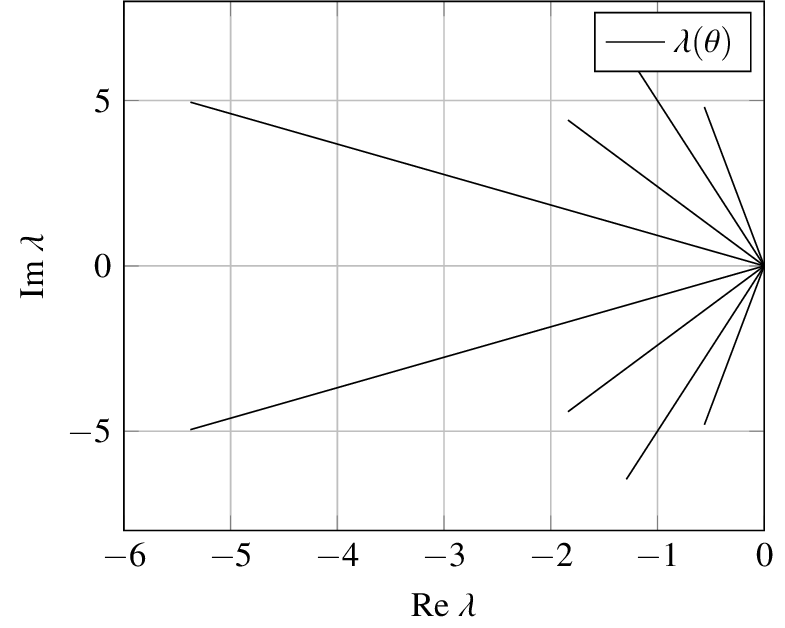} 
\caption{The smallest branch points at $M=10$ for the quantum wire model
  (circles) and paths $(1-\theta)\lambda_*$ used for determining which
  branches $E_n(\lambda)$ meet at $\lambda_*$. See also
  Figure~\ref{fig:qdot-multiple-points2}.\label{fig:qdot-multiple-points}}
\end{center}
\end{figure}

\begin{figure}[ht]
\begin{center}
\myinputfig{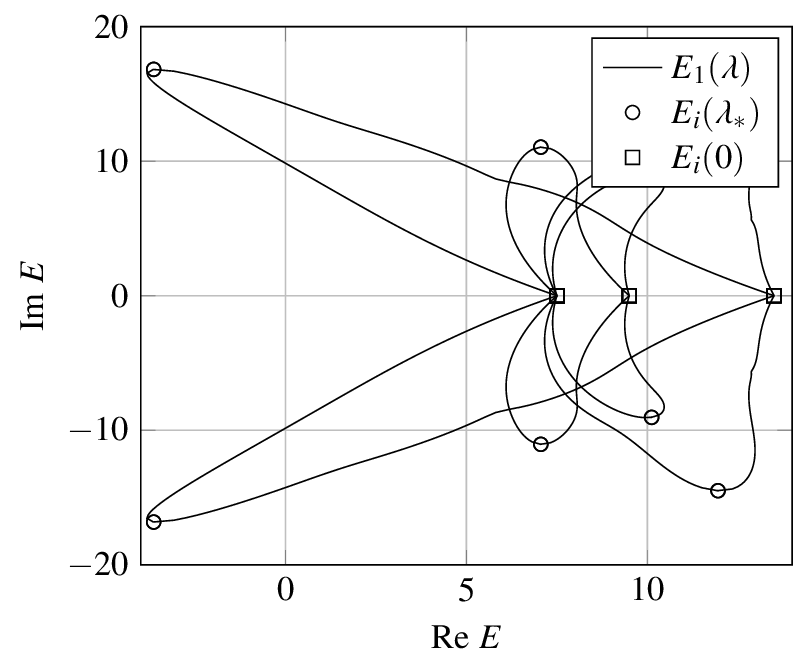} 
\caption{The eigenvalue $E_0(\lambda_*)$ (squares) corresponding to
  the branch points $\lambda_*$ in
  Figure~\ref{fig:qdot-multiple-points}. Also shown are the paths
  $E_n[(1-\theta)\lambda_*]$ for the branching eigenvalues, determining
  which branches $E_n(\lambda)$ actually meet at $\lambda_*$.\label{fig:qdot-multiple-points2}}
\end{center}
\end{figure}

To illustrate the continuation procedure for determining which
eigenvalues branch a given $\lambda_*$, we have shown a number of the
branch points involving the ground state $E_0(\lambda)$ and some other
$E_n(\lambda)$ in Figure~\ref{fig:qdot-multiple-points}. We construct a
path $\lambda(\theta) = (1-\theta)\lambda_*$ (also shown) and
compute the eigenvalues of $H(\lambda(\theta))$ that branch at $\lambda_*$.
As these are continued to $\lambda(1)=0$, they will be equal to
unperturbed energies, and the branches are easy to determine. In
Fig.~\ref{fig:qdot-multiple-points2} the eigenvalue paths
$E_0(\lambda(\theta))$ and $E_n(\lambda(\theta))$ are shown. It is clearly seen
that the ground state and some excited state meet at
$\lambda_*$. This also shows how the continuation procedure may be
used to find the non-dominant singularities involving the ground
state, which can be taken as input for approximant construction.

\subsection{A helium-like model}
\label{sec:helium}

The final example is a helium-like model in one spatial dimension, also
considered by Herman and Hagedorn \cite{Herman2009}. It shares many of
the qualitative features with the true helium atom in three spatial
dimensions, and our goal is to determine the ROC for a Møller--Plesset
calculation of the ground state energy. The model has the Hamiltonian
\begin{eqnarray*}
  \mathcal{H} &=& \sum_{i=1}^2 \left(
    -\frac{1}{2}\frac{\partial^2}{\partial x_i^2} -
    Z\delta(x_i)\right) + \delta(x_1-x_2) \notag \\ &=& \mathcal{H}_0 +
   \mathcal{V},
\end{eqnarray*}
where $\mathcal{V}=\delta(x_1-x_2)$ is the inter-electron interaction.
The two electrons also interact with the nucleus of charge $Z$ via
$\delta$-function potentials. We set $Z =
1.38$ for the calculations.

For Møller--Plesset perturbation theory we rewrite the
physical Hamiltonian as
\begin{equation}
  \mathcal{H} = \mathcal{H}_0 + \mathcal{U}^\text{HF} + (\mathcal{V} - \mathcal{U}^\text{HF}),
\end{equation}
where the Hartree--Fock operator $\mathcal{U}^\text{HF}$ is defined
in the usual way \cite{Helgaker2002}. Since we are going to treat
$\mathcal{V}-\mathcal{U}^\text{HF}$ as a perturbation, we introduce a parameter
$\lambda$, viz,
\begin{equation*}
  \mathcal{H}^\text{HF}(\lambda) = \mathcal{H}_0 + \mathcal{U}^\text{HF} +
  \lambda (\mathcal{V} - \mathcal{U}^\text{HF}).
\end{equation*}
for which $\mathcal{H}^\text{HF}(1) = \mathcal{H}$.  We now wish to
determine whether the MP series converges, \emph{i.e.}, the radius of convergence must
be greater than $R=1$.

For this, we need to determine the Hartree--Fock basis and
energies. We do this using a linear finite element basis over the
interval $[-L,L]$ using $n$ sub-intervals of length $\Delta x = 2L/n$
to obtain the usual Roothan--Hall equations which are solved
iteratively \cite{Helgaker2002}. In our calculations we have $L=15$
and $n = 1000$, giving $\Delta x = 0.03$.  The exact solution to the
Hartree--Fock ground state can be obtained in this case \cite{Herman2009} which provides a good check on the accuracy of the
implementation.

The discretization parameters are $\Delta x$, $L$ and the
number of single-particle functions $M$ we use in the MP series. However, we
will consider the spatial discretization parameters to be fixed
and sufficient for an ``exact'' treatment of the one-body
Hamiltonians, and instead only focus on $M$.

The ground state is a singlet state, for which the spatial
wave function is symmetric with respect to interchange of $x_1$ and
$x_2$,  and has positive parity, \emph{i.e.},
\begin{equation*}
  \psi(x_1,x_2) = \psi(x_2,x_1) = \psi(-x_1,-x_2). 
\end{equation*}
There are otherwise no dynamical symmetries in this problem.

\begin{figure}[ht]
\begin{center}
\myinputfig{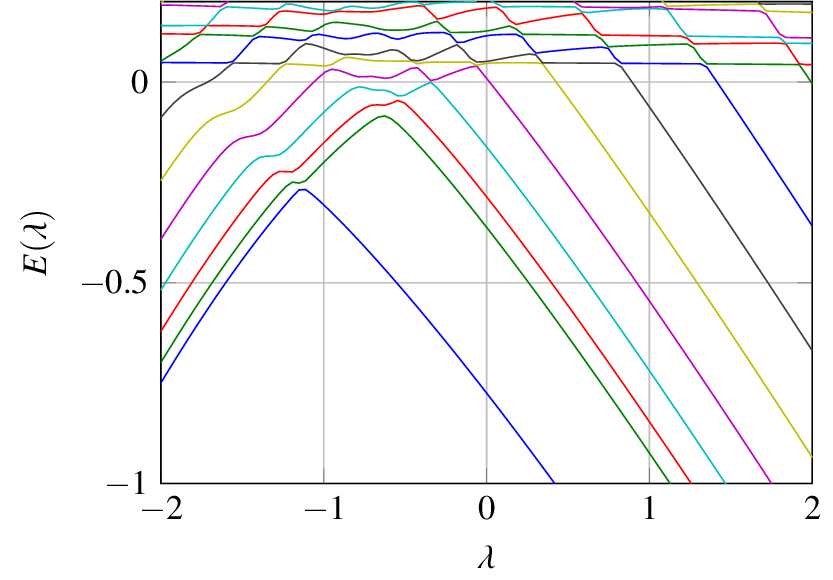}
\caption{(Color version online) Plot of the eigenvalues $E_n(\lambda)$ for the helium-like
  model, using $M=42$ single-particle functions. Note crossing around
  $\lambda\approx -1.1$ with small imaginary part. We therefore expect
  interesting crossings with $E_1$ and some other $E_n$ to have
  significant imaginary part since no other avoided crossings
  involving the ground state is visible.\label{fig:helium-lambda-sweep}}
\end{center}
\end{figure}

Figure~\ref{fig:helium-lambda-sweep} shows a parameter sweep of the
eigenvalues of $\mathcal{H}^\text{HF}(\lambda)$. In this case, it is
clear that there is a back-door intruder around $\Re\lambda_*\approx
-1.1$, so $R \lesssim 1.1$ is obvious. It is highly likely that it corresponds
to a $\beta$ type singularity in the exact problem, as there are
clearly several close crossings (verified in our computations) with
continuum states (with positive energy at $\lambda=0$) near
$\lambda_*$. This is also supported by Herman and Hagedorn \cite{Herman2009}. 

However, it is not clear whether or not there are
\emph{other} branch points with, say, a small real part and imaginary
part $\Im\lambda_* = 0.7$, which would imply $R<1$ and hence an
ultimate \emph{divergence} of the MP series.

The position or existence $\beta$ singularities in
the discrete problem is highly dependent on the basis chosen \cite{Olsen1996,Sergeev2005}, so our conclusions must be taken
relative to our discretization, which does include ``diffuse'' basis
functions approximating the continuous spectrum.

\begin{figure}[ht]
\begin{center}
\myinputfig{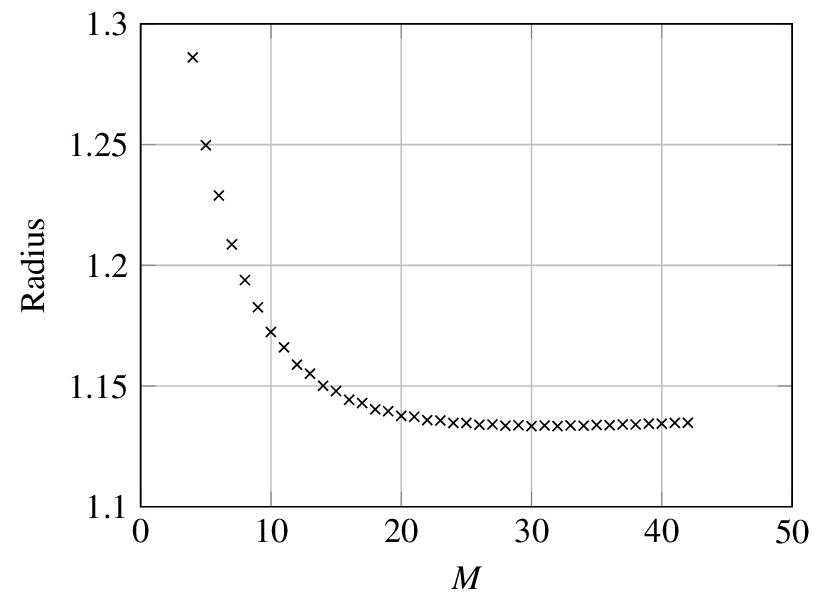}
\caption{Convergence radius of the helium-like model as function of the
  number of single-particle functions $M$. A rapid convergence
  towards $R\gtrsim 1.13$ is seen.\label{fig:helium-ROC}}
\end{center}
\end{figure}

\begin{figure}[ht]
\begin{center}
\myinputfig{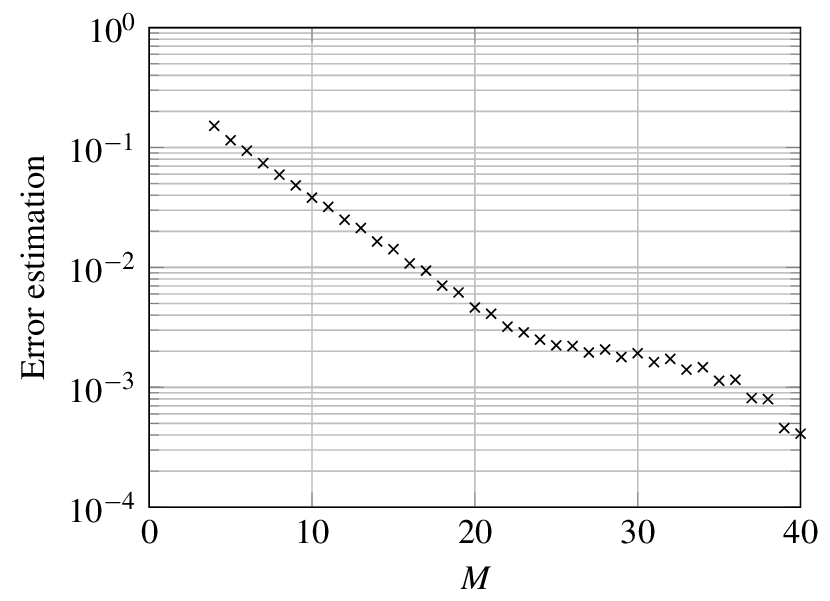}
\caption{Estimated error in the ROC for the helium-like model as
  function of the number of single-particle functions $M$. An
  exponential error is observed, \emph{i.e.}, $|R(M)-R(\infty)|\sim
  \exp(-\beta M)$.\label{fig:helium-ROC-error}}
\end{center}
\end{figure}

We compute, like in the preceding numerical examples, the
dominant branch point $\lambda_{*,0}$ for various $M$ and study its
behavior. For $M\le 13$ we used the
Arnoldi method, while for $M>13$ the inverse iteration method.
It turns out that, in fact, $R > 1$; the avoided crossing
does indeed come from the dominant branch point. In
Figure~\ref{fig:helium-ROC} the ROC as function of $M$ is
shown. Clearly, it 
stabilizes around some value $R \gtrsim 1.12$. In Figure~\ref{fig:helium-ROC-error} an error estimate is plotted, based on the
largest $M=42$,  and clearly the error behaves like $\exp(-\beta M)$
where $\beta>0$ is a constant. 

\begin{figure}[ht]
\begin{center}
    \myinputfig{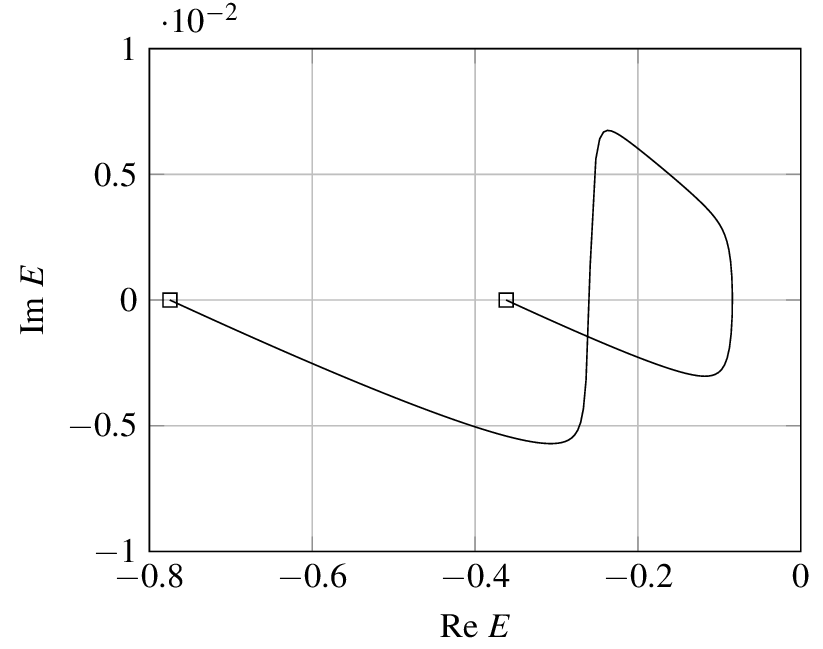}
\caption{Eigenvalue path for $M=42$ for the dominant branch point
  $\lambda_{*,0}$. As $\lambda$ takes values along the path
  $(1-\theta)\lambda_*$, the branching eigenvalues start at
  $E_n(\lambda_*)=E_{n'}(\lambda_*)$ and moves to $E_n(0)$ and
  $E_{n'}(0)$, so that which branches meet can be determined. Here,
  $n=0$ and $n'=1$.\label{fig:helium-lambda-path}}
\end{center}
\end{figure}

In Figure~\ref{fig:helium-lambda-path} the paths $E_0(\lambda(\theta))$ and
$E_1(\lambda(\theta))$ are shown, where $\lambda(\theta) =
(1-\theta)\lambda_{*,0}$. This corresponds to
Figure~\ref{fig:qdot-multiple-points2} for the quantum wire model. The
path is surprisingly complicated, showing that the eigenvalues may
take complicated deviations from their initial unperturbed values as
$\lambda$ varies in a straight line.

\section{Conclusion}
\label{sec:conclusion}

We have described a numerical procedure to determine the singularities
of the eigenvalues $E_n(\lambda)$ of $H_0 + \lambda V$. Using a
continuation technique that tracks eigenvalues as function of $\lambda$,
the dominant singularity can be found. A simple generalization of this
will enable the classification of other singularities with respect to
which eigenvalues branch at $\lambda_*$.  By continuing Step~2 and Step~3
in Algorithm~\ref{alg:roc} after $\lambda_{*,0}$ has been found, one
can find the secondary dominating branch point and so on. 

The method has been successfully applied to instructive examples, and
in particular the convergence of a Møller--Plesset perturbation series
for a helium-like model was established.

The most important virtue of the method is that it searches the whole
complex plane for singularities, and also in principle can find
\emph{all} these. This allows a much more detailed mapping of the
singularity structure than the standard methods based on the
asymptotic form of the terms in the series.

Computing the singularities of $E_n(\lambda)$ is much harder than
computing only $E_n(\lambda_\text{phys})$. One cannot hope to be able
to compute the whole set of singularities for a very large many-body
system.  However, obtaining insight into the distribution of
singularities for ``typical'' quantum systems, such as the examples
considered in this paper and others \cite{Sergeev2005}, makes the
construction and analysis of general resummation schemes
easier \cite{Katz1962,Goodson2003}. Considering that popular approaches
to the many-body problem such as coupled cluster methods can be viewed
in terms of perturbation series only serves to emphasize the
importance of calculations of singularity structures.

\section*{Acknowledgments}

This work is supported by the Norwegian Research Council.
This article presents results of the Belgian 
Programme on Interuniversity Poles of Attraction, initiated by the
Belgian State, Prime Minister's
Office for Science,
Technology and Culture, the Optimization in
Engineering Centre OPTEC of the K.U.Leuven, and the project STRT1-09/33 
of the K.U.Leuven Research Foundation.

\end{document}